\documentclass[10pt]{article}
\usepackage[T1]{fontenc}
\usepackage[utf8]{inputenc}
\DeclareUnicodeCharacter{2009}{\,}
\DeclareUnicodeCharacter{00A0}{~}
\usepackage[a4paper, total={6in, 8in}]{geometry}

\usepackage{xcolor}
\definecolor{purple}{HTML}{B8008A}

\newcommand{\edits}[1]{\color{black} #1 \color{black}}

\newcommand{\camilo}[1]{\color{orange} #1 \color{black}}

\usepackage{graphicx, array, blindtext}
\usepackage{subfig}
\usepackage[export]{adjustbox}
\usepackage{enumitem}

\usepackage{gensymb}
\usepackage{mathtools}
\usepackage{amsthm}

\usepackage{longtable}
\usepackage{authblk}

\begin{document}

\title{Atomistic modeling of the hygromechanical properties of amorphous Polyamide 6,6}

\author[1]{Karim Gadelrab~\footnote{Corresponding author, karim.gadelrab@us.bosch.com}}
\affil[1]{ Research and Technology Center, Robert Bosch LLC, Watertown, MA 02472, USA}
\author[2]{Armin Kech}
\affil[2]{Component Design, Reliability and Validation Polymers, Corporate Sector Research and Advance Engineering, Robert Bosch GmbH, Robert-Bosch-Campus 1, 71272 Renningen, Germany}
\author[2,3]{Camilo Cruz}
\affil[3]{Clément Ader Institute, IMT Mines Albi, Campus Jarlard, 81013 Albi, France}
\date{}     
\setcounter{Maxaffil}{0}
\renewcommand\Affilfont{\itshape\small}

\maketitle

\begin{abstract}

Polyamide 6,6 (PA66) is a key engineering polymer whose unique mechanical properties arise from strong interchain hydrogen bonding. However, its hygroscopic nature makes it highly sensitive to water uptake, which markedly alters its thermomechanical behavior. {Contrary to traditional experimental approaches,} this study uses atomistic molecular dynamics (MD) simulations to investigate the role of water in modifying the glass transition temperature ($T_g$) and the {viscoelastic response} of amorphous PA66. Simulations capture a non-monotonic dependence of $T_g$ on water content. At low water concentrations, isolated water molecules bind to amide groups and restrict chain mobility, while beyond $\sim 2.5$~wt.\% water clustering disrupts the hydrogen-bond network and causes a pronounced $T_g$ depression. Analysis of amide-group fluctuations reveals a master correlation between local segmental dynamics and bulk density, {verifying the known temperature-humidity equivalence in terms of density variation.}
The computed Young’s modulus exhibits systematic softening with increasing temperature and water content, consistent with experimental trends, albeit a more pronounced impact of water at low temperatures. Time-temperature superposition behavior is
observed for both dry and hydrated systems.
This work provides molecular-scale information on the hygromechanical coupling in PA66 and demonstrates the ability of MD simulations to predict water-induced transitions that govern the macroscopic behavior of polyamides.

\end{abstract}

\section{Introduction}

The unique chemistry of polyamides makes them one of the most relevant engineering polymer families that are used in load-bearing applications, especially mechanical components\cite{Ehrenstein,Feldman03042017}. The hydrogen bonding in the amide groups results in a strong interaction between the chains, which is reflected in the relatively high stiffness and yield strength \cite{holmes1955crystal}. Unfortunately, the same polar nature of polyamides makes them hygroscopic, exhibiting a pronounced tendency to moisture uptake \cite{puffr1967structure,reuvers2015plasticization,venoor2021understanding}. After adsorption, water molecules diffuse readily into the amorphous phase of the polyamide matrix by different mechanisms. Water molecules occupy the free volume of the amorphous phase \cite{cohen1959molecular,kumins1968free}, form hydrogen bonds with amide groups that disrupt the configuration of the polymer matrix \cite{merdas2002factors,moy1980epoxy} and can form clusters after a certain level of water uptake \cite{sabard2012influence,le2014water}. \edits{The absorption of water affects the properties of polyamide materials in different ways \cite{antiplas02}. Typically occurring at low temperatures or low water concentrations, an antiplasticization phenomenon takes place. In this state, water molecules occupy existing cavities within the amorphous phase rather than expanding it. This process results in a macroscopic stiffening of the material due to the steric hindrance provided by "caged" or immobile water molecules \cite{antiplast99,Kehrer2024}. However, at elevated temperatures and concentrations above a critical threshold, water acts as a plasticizer of the polymer matrix, which can lead to a significant reduction in stiffness (decrease in Young’s modulus) and increased ductility (higher elongations at rupture in a tensile test) \cite{miri2009effect,polym15163387}. In this case,} the presence of water in the amorphous phase facilitates the cooperative motion of chain segments and induces a reduction in $T_g$. Dry PA66 with crystal contents ranging between $35-45$~wt.\% exhibits a $T_g$ close to $90$\textdegree C \cite{Ehrenstein}. Under $100$\% relative humidity, $T_g$ drops below $0$\textdegree C \cite{broudin2015water,parodi2018prediction,arhant2016modelling, launay2013modelling}, which highlights the strong sensitivity of PA66 to moisture uptake. Depending on the level of water uptake, the absorption of water may cause some regions or a whole polyamide component to transition from a glassy to a rubbery state under typical service temperatures. In consequence, water uptake in polyamide parts could affect their mechanical integrity, dimensional stability, and long-term durability. This underscores the importance of accounting for environmental moisture in the design and dimensioning of engineering components made of polyamide-based materials.

The mechanical response of polyamide materials is naturally influenced by temperature, which governs both molecular mobility and activation of deformation mechanisms \cite{launay2013modelling}. As temperature increases, semi-crystalline thermoplastics generally exhibit enhanced ductility and reduced strength \cite{khan2006thermo,rozanski2013plastic}. These changes are especially pronounced in the amorphous regions, where large-scale molecular rearrangements and segmental motion control different transitions and the evolution of material properties. 
The mechanical behavior of polyamides is not dictated solely by temperature; it also depends on the interplay with deformation rate, which is a manifestation of the viscoelastic nature of the amorphous phase in those semi-crystalline materials \cite{fabre2018time,billon2012new, maurel2015thermo}. 
Interestingly, the mechanical signature driven by temperature closely mimics the one induced by moisture uptake. This is a consequence of the main role of the amorphous phase during the deformation of semi-crystalline polyamide matrix \cite{miri2009effect,launay2013modelling}. Water, acting as a plasticizer, facilitates local molecular mobility in the amorphous phase, depresses the glass transition temperature and alters chain relaxation behavior. This coupling of temperature, water content, and strain rate poses significant challenges in the development of predictive physics-based mechanical constitutive models for polyamides\camilo{\cite{launay2013modelling,Engelhardt13,SharmaP2020,Dyck2024}.} 

From an industrial perspective, the release of polyamide-based engineering components requires extensive and costly physical testing to account for the impact of humidity in their mechanical performance. Those tests typically involve component conditioning under different humidity environments during times going from a couple of weeks up to several months, depending on the technical requirements. In this context, robust virtual dimensioning of polyamide-based components considering water uptake is a key lever for reducing development time and exploring more extensively the design spaces. State-of-the-art simulation-based mechanical design of thermoplastic components makes use of finite element analysis coupled with homogenization methods (accounting for the presence of reinforcement fibers, for example). For running those simulation techniques, one needs to define pertinent constitutive laws of the involved phases. In the case of polyamide components, we need to identify mechanical constitutive models that emulate correctly the dependence on water content. In addition to standard identification protocols based on physical testing at the coupon level, we can imagine a full in-silico approach based on atomistic simulations. MD simulations constitute a powerful tool for understanding the interaction between polyamide chains and water molecules and, finally, predicting changes in material properties such as glass transition temperature and elastic modulus. The vision is that atomistic simulations become a reliable instrument for estimating the mechanical behavior of a semi-crystalline polyamide system with any water content. 

\edits{Several research works have used atomistic simulations for investigating polyamide-water systems. In their premier work, Goudeau et al. \cite{Goudeau1} studied via atomistic MD simulations how water molecules alters the local structure and volumetric properties of hydrated amorphous PA66. They verified in-silico the different regimes of water distribution within amorphous PA66 as function of water concentration, going from tightly bounded water up to loosely bounded and water clusters (formed preferably at chain ends). Authors showed also that this water organization correlates directly with the transition between antiplasticization (hole filling by tightly bounded water) and plasticization (net increase of free volume by presence of loosely bounded water) behavior. These features of water organization have been also successfully correlated with the changes in the distribution of the hydrogen bonds driven by the presence of water in nanoconfined PA66 \cite{Eslami}. MD simulations allowed also to study the dynamics of water molecules and PA66 chain segments \cite{Goudeau2}. The analysis revealed that water molecules exist in bimodal dynamical states, characterized by differing reorientation times and hopping frequencies between cavities. Ultimately, the work demonstrated that water's impact on polyamide dynamics is a complex interplay of localized hydration and temperature-dependent diffusion mechanisms. Ikeshima et al. \cite{IkeshimaDaiki2019Dmop} generated a semi-crystalline MD cell for studying the effect of water on the deformation behavior of PA6. The work explained the reduction of Young's modulus and yield stress in the hydrated system (0.75 wt.\%) by the reduction of the non-bonded potential energy between pairs of amide functional groups. Water would disturb the internal cohesive energy of the PA6 and would act as a lubricant during system deformation. More recently, Pilz et al. \cite{pilz2023calculation} used MD simulations of hydrated amorphous PA66 and $\alpha$-crystalline PA66 cells for estimating the effective mechanical properties of the hydrated semi-crystalline system by using homogenization schemes. Despite showing a reduction of Young's modulus with increasing water content in amorphous PA66, the proposed models underestimate the softening effects measured experimentally. This last result evidences the need of understanding and quantifying with more accuracy via atomistic simulations the impact of water on the mechanical behavior of amorphous polyamide systems.} 

In this work, we utilized MD simulations to investigate further the impact of water on the internal structure of amorphous PA66 and discussed the effect of water as a plasticizing agent. We computed water-dependent $T_g$ values from amorphous polymer bulk simulations. The segmental fluctuations were quantified and discussed in the context of the change of polymer density. Finally, we estimated Young's modulus of different amorphous PA66/water systems, taking into account the role the deformation rate plays in the viscoelastic response of those systems. This investigation covered a wide range of temperatures and water content. \edits{Besides an alternative in-silico validation of the antiplasticization-plasticization transition at low water content, this research work makes two main original contributions to the field: 1) The explanation of the temperature-humidity equivalence for volumetric properties (density) in terms of amide group fluctuations, and 2) the MD evidence that the time-temperature superposition principle for viscoelastic behavior also holds for amorphous PA66 containing water molecules. This contribution particularly paves the way towards more accurate atomistic simulations of the mechanical behavior of humid amorphous PA66 and sheds light} on the development of advanced algorithms that account for the crystalline phase within semi-crystalline engineering thermoplastics.


\section{{Simulation models and Methods}}


We used an MD bulk simulation cell to investigate the thermo-mechanical behavior of amorphous PA66 under dry and humid conditions. The material system consisted of 12 polymer chains, each composed of 80 monomers, which were initially packed at low density using the MedeA software package \cite{thompson2022lammps} with periodic boundary conditions in the three directions.This simulation model corresponds to a PA66 with a uniform molecular weight distribution equal to 18.1 kg/mol. Note that for amorphous PA66 the entanglement molecular weight $M_e$ is $\sim$ 2 kg/mol and the critical molar mass $M_n^c$, determining the brittle-ductile transition, oscillates between $3M_e$ and $5M_e$ \cite{lottier:tel-04172538}. 

Water molecules were introduced into the amorphous PA66 matrix to set up humid systems with different water contents by mass $C_w$. Water content by mass is defined as follows: $C_w = m_w/(m_w+m_p)$, where $m_w$ refers to the total water mass and $m_p$ is the \edits{dry} PA66 mass. The system with the highest water content was set to $C_w$ = 8.7 wt.\%. Independently of the exposure temperature, semi-crystalline PA66 systems can absorb water up to $\sim8.8$~wt.\%  \cite{parodi2018prediction,ledieu:pastel-00547111}. By assuming that water molecules cannot penetrate the crystalline phase, we can expect in fact higher water concentrations in the amorphous phase up to $\sim11.9$ wt.\% (supposing $26$ wt.\% of crystal content). 

\edits{Chain connectivity and force field parameters followed the PCFF (Polymer Consistent Force Field) functional form \cite{doi:10.1021/jp980939v}, where bonded interactions include anharmonic bond stretching, angle bending, torsions, and cross-coupling terms (bond–bond, bond–angle, angle–angle), while non-bonded interactions are represented by a 9-6 Lennard-Jones potential and fixed partial charges for electrostatics. PCFF was parameterized against a combination of experimental data and quantum calculations to reproduce polymer properties such as densities, conformational energetics, and vibrational properties. It is worth noting that PCFF, similar to other classical force fields, has its own limitations including the absence of explicit polarization, lack of descriptive power of reactivity and bond breaking, and potential transferability issues outside the chemical space for which it was originally parameterized. Nonetheless, the descriptive power of PCFF as a class II force field \cite{rigby2004fluid} makes it ideal for the accurate representation of intermolecular and intramolecular interactions for PA66/water system investigated in this study}. Finally, The Large-scale Atomic/Molecular Massively Parallel Simulator (LAMMPS) was used for all MD simulations \cite{LAMMPS}.

To achieve a well-equilibrated polymeric system with appropriate density and structure, a multi-step equilibration protocol was implemented. Initially, the temperature was regulated using the Nose-Hoover thermostat with a timestep of 0.3 femtoseconds (fs). The equilibration process began with an NVT ensemble \edits{(i.e. fixed number of particles, volume, and temperature)} of packed nylon/water molecules with low density at a fixed temperature of $800$ K to facilitate polymer mobility and effective chain relaxation. Subsequently, the system was subjected to an NPT ensemble \edits{(i.e. fixed number of particles, pressure, and temperature)} at $600$ K for 4 ns to further relax the structure and achieve a stable density within a cubic simulation cell.

\subsection{Assessment of glass transition temperature}
\label{sec:MD-Tg}

The glass transition temperature $T_g$ is a key physical parameter of polymers, which marks the transition between the glassy and rubbery states of the amorphous phase. The experimental methods to determine $T_g$ in thermoplastic polymers take advantage of the change in various material properties (mechanical, dielectric, or thermal) through the glass transition. For example, $T_g$ can be determined by techniques such as Dynamic Mechanical Analysis (DMA) \cite{fabre2018time}, dielectric measurements\cite{hensel1996temperature}, or Differential Scanning Calorimetry (DSC) \cite{humeau2018influence}. In addition to those methods, $T_g$ can also be identified using the change in thermal expansion coefficient between the glassy and rubbery states. In principle, the density curve as a function of temperature $\rho(T)$ should exhibit a clear change in slope at the glass transition. In this work, we propose to estimate via MD simulations the signal of $\rho(T)$ to identify the glass transition temperature of the different material systems. 

More specifically,  we performed a series of independent NPT simulations in a temperature range from $420$ K to $230$ K sampled at a step size of $10$ K (each simulation runs for $5$ ns). The density of the system was quantified independently for each single NPT simulation. Each lower-temperature simulation was initiated using the final configuration of the preceding higher-temperature run. This stepwise approach looks to minimize the likelihood of frozen or trapped states that could arise from excessively continuous rapid cooling by thermalizing the system at the beginning of each run. \edits{This approach is independently followed for every $C_w$; hence, cross correlation between different PA66/water simulations is avoided.} The estimation of $T_g$ is based on the observation that density varies linearly with temperature in the glassy (low-temperature) and rubbery (high-temperature) regimes. $T_g$ is conventionally identified by fitting linear trends to these two regions and determining the temperature at which they intersect.

\subsection{Assessment of Young's modulus}

The temperature quasi-equilibrated structures obtained for evaluating the density in the previous section served as a starting point for the simulations of mechanical deformation. To evaluate the mechanical properties of the equilibrated system, uniaxial deformation simulations were performed under constant strain-rate conditions. Stress-strain responses were obtained by applying controlled tensile and compressive deformation along a single axis. This approach mitigates the formation of cavities when large deformations are reached and looks for emulating the constraint-free deformation of the sample's cross-section in a standard tensile test. The Young’s modulus of the material was determined from the linear regime of the stress-strain curve in the small strain range ($|\epsilon|<$  3\%). All MD structures (dry and wet amorphous PA66) exhibited isotropic elastic behavior, given the fact that the average difference between the identified Young’s modulus values in the x, y and z directions were lower than $0.05$ GPa for a given temperature and strain rate. Therefore, only data for the x direction are shown in the next section. Five different strain rates, ranging from 10$^{10}$ to 10$^{8}$~1/s, were used to assess the rate-dependent mechanical behavior of the various PA66 systems.

\section{Results and Discussion}
\subsection{Effect of water on Tg}
\label{sec:results-Tg}


The glass transition is a second-order-like transition that occurs in a temperature range in which the molecular mobility of the polymer chains changes substantially. The definition of a singular $T_g$ as a characteristic function of the material depends strictly on the measurement kinetic conditions \cite{MarkJamesEdward2007PPoP,HalaryJean-LouisMdmp} and a particular convention of data post-processing. Compared to experimental {dilatometric} measurements, the glass transition observed in MD simulations spans a larger temperature range, often extending over tens of degrees. This broadening effect is hypothesized to arise due to finite-size effects in simulations, the limited timescale available for structural relaxation, and inherently rapid cooling rates \cite{patrone2016uncertainty}. Furthermore, the presence of water molecules {impacts}{ the free volume in the polymer system,} facilitates molecular mobility, and further widens the transition range. The challenge in determining $T_g$ also stems from statistical fluctuations in density data, particularly at extreme temperatures, where linear extrapolation can be highly sensitive to noise \cite{patrone2016uncertainty,lukasheva2017influence}. This sensitivity requires careful analysis and the averaging of multiple independent simulations to obtain a reliable {glassy baseline.}

Patrone et al. \cite{patrone2016uncertainty} presented a hyperbola model to estimate $T_g$ from the density plot, with the advantage of fitting the entire density profile and removing the subjectivity of the double-tangent method. As will be shown in the following, the hyperbole model proves to be a versatile function that can extrapolate between two linear regimes. The hyperbola expression is 

\begin{equation} \label{eq:hyperbola}
    \rho(T)=\rho_{0}-a\left(T-T_{0}\right)-b\left[\frac{1}{2}\left(T-T_{0}\right)+\sqrt{\frac{\left(T-T_{0}\right)^{2}}{4}+e^{c}}\right]
\end{equation}

Here, $\rho(T)$ is the temperature-dependent density. $T_0$, $a$, $b$, and $c$ are fitting constants. The parameter $c$ smoothens the transition between the asymptotic limits of the function. The discontinuity in the slope occurs at $T_0$, recovering the double-tangent method. 

\begin{figure}
    \centering
    \includegraphics[height=0.55\textwidth]{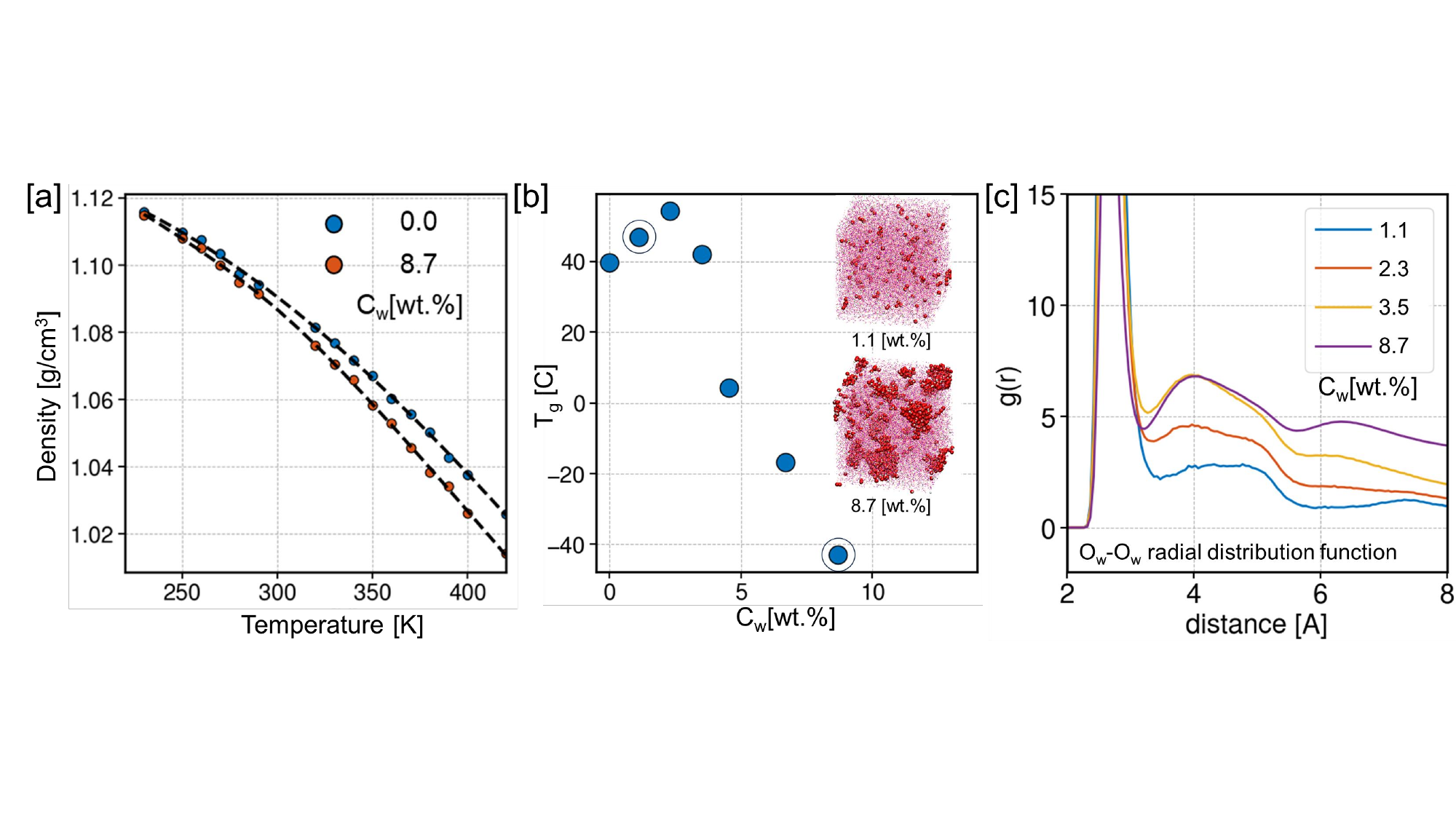}
    
    \caption{
    MD analysis of the glass transition behavior of amorphous PA66 as a function of water content. (a) Density $\rho$ as function of temperature $T$ for the dry system ($C_w = 0$~wt.\%) and a hydrated system ($C_w = 8.7$~wt.\%), illustrating the thermal expansion and the density reduction associated with water-induced swelling. Dashed lines are fits obtained using Eq.~\ref{eq:hyperbola} (b) Variation of $T_g$ with $C_w$. An increase in $T_g$ is observed at low water contents ($C_w\leq 2.5$~wt.\%), suggesting a binding effect of isolated water molecules (H$_2$O MD snapshot at $C_w = 1.1$~wt.\%) that locally stabilizes the amide hydrogen-bond network. Beyond this concentration, $T_g$ decreases continuously as water clustering (H$_2$O MD snapshot at $C_w = 8.7$~wt.\%) disrupts inter-chain hydrogen bonds and promotes segmental mobility, leading to an effective softening of the amorphous matrix.  
    (c) Radial distribution function $g(r)$ of water oxygen atoms at various water contents. A peak at 2.8~{\AA} corresponds to the nearest-neighbor O-O distance, while the emergence of a secondary peak near 4~{\AA} (fully developed at $C_w = 3.5$~wt.\%) and a third peak around 6.5~{\AA} indicates progressive clustering of water molecules. These features mark the transition from well-dispersed water at low concentrations to the formation of bulk-like water clusters at higher $C_w$, consistent with the observed $T_g$ dependency on water content.  
        }

    \label{fig:rdf}
\end{figure}

Fig.~\ref{fig:rdf}a shows the $\rho(T)$ plot {using the MD protocol described in §\ref{sec:MD-Tg}, which emulates an effective cooling rate of $\sim 10^{11}$ K/min}. 
Only data points for the dry system ($C_w = 0.0$~wt.\%) and one wet system ($C_w = 8.7$~wt.\%) are shown for clarity. {Results show the expected reduction of $\rho$ by increasing temperature (capturing the phenomenon of thermal expansion), as well as the decay of system's density with the increase of water content. This last effect corresponds to the volumetric expansion of the amorphous PA66 driven by the water uptake. In other words, MD can reproduce the phenomenon of polymer swelling.} \edits{Based on our MD results, the linear coefficient of thermal expansion of the dry amorphous PA66 at room temperature comes out to be 1.34$\times10^{-4}$ 1/K, which is in agreement with an experimental reference value of 8.7$\times10^{-5}$ 1/K, measured on an injection-molded semi-crystalline PA66 \cite{polym15163387}. Deviation is mainly  related to the presence of a crystalline phase in the actual measured samples.} 

It is observed that within the temperature range investigated in this work ($230$~K-$420$~K), no clear asymptotic behavior in density is observed especially at edge temperatures. This problem might be solved by increasing the temperature limits, but that adds lever uncertainty to the exact point of slope change, especially finding a {pseudolinear} regime at {hundreds} of degrees above the transition point. Moreover, the presence of mobile species (water) in the polymer matrix introduces significantly more relaxation timescales which broadens the transition region. However, the hyperbolic function expression in Eq.~\ref{eq:hyperbola} managed to fit the $\rho(T)$ data quite well, showing a continuously curved shape over the investigated temperature range. {The fitting process allows to identify $T_0$ in Eq. \ref{eq:hyperbola}, which we assume corresponds to $T_g$ under the conditions of the MD protocol. The identified $T_g$ as a function of the water content} is plotted in Fig.~\ref{fig:rdf}b. 
 
The MD results captured the overall reduction in $T_g$ with the increase of water concentration in agreement with various experimental reports \cite{broudin2015water,parodi2018prediction,arhant2016modelling, launay2013modelling}. {We estimated a $T_g$ of $39.5^o$C for dry amorphous PA66 and a drop of $T_g$ to $-42.8^o$C for a system with $8.7$~wt.\% of water. \edits{It represents on average a reduction of $\sim9.7^o$C per \% by weight of water in amorphous PA66. For comparison, using dynamic mechanical analysis (heating rate of $1^o$C/min) on a PA66 ($16.5$~kg/mol, $35$ \% crystal content by weight), Broudin et al. \cite{broudin2015water} measured a $T_g$ of $58 \pm 2^o$C in the dry state and a $T_g$ of $\sim4^o$C for a water content of $9.6$ \% by weight in the amorphous phase. This corresponds on average to a drop of $\sim5.7 ^oC$ per \% by weight of water in amorphous PA66}. A similar variation of $T_g$ with water content has been reported for a PA66 ($35.6$~wt.\% crystal content) reinforced with $35$~wt.\% glass fibers \cite{launay2013modelling}.} 
Such differences can be rationalized by {the confinement of the amorphous phase in an actual semi-crystalline PA66 system,} the inherent limitations of MD in terms of system size, and degree of polymerization. 

However, it is notable that semi-quantitative values of $T_g$, which span a wide temperature range, can be obtained from MD. Contrary to some phenomenological models \cite{MarkJamesEdward2007PPoP}, the change in $T_g$ with the water content is not a monotonous decreasing function, in fact, MD simulations show a moderate increase in $T_g$ at water concentrations lower than $2.5$~wt.\%. At higher water concentrations, we observe a classical downward trend of $T_g$ with increasing water content. This result points to a binding role for water in the polymer matrix at low water concentrations \cite{Goudeau1}. Well-dispersed water molecules would interact strongly with the amide groups and limit their mobility.
Only when clusters of water molecules form \cite{gaudichetmaurin:pastel-00001528}, the softening of amorphous PA66 takes place. 

A visualization of the distribution of water molecules is shown in Fig.~\ref{fig:rdf}b at low ($1.1$~wt.\%) and high ($8.7$~wt.\%) water content. Water is well-dispersed in the polymer matrix at low $C_w$, but forms large clusters at high $C_w$. {A more quantitative description of the water distribution} can be obtained by analyzing the water oxygen-oxygen radial distribution function $g(r)$ for different values of $C_w$ (see Fig.\ref{fig:rdf}c). The first peak at $2.8$~\AA corresponds to the O-O distance between two neighboring water molecules. Additional layers of water molecules (clustering) should exhibit peaks at larger distances. 

At $C_w = 3.5$~wt.\%, a fully developed peak can be observed at $4$~\AA~and traces of a third peak at $6.5$~\AA~start to build up. These peaks indicate the transition of {uni-molecular water distribution} to water clusters that closely match the $g(r)$ of bulk water \cite{wang2011density} at values of $C_w > 3.5$~wt.\%, which correlates well with the persistent drop of $T_g$. {The critical water content of $2.5$~wt.\% in amorphous PA66 would correspond to $\sim 1.8$~wt.\% in a semi-crystalline system (supposing a crystallinity of $35$~wt.\%).} 

This antiplasticization effect at lower water contents in polyamides was observed \edits{in the early eighties} by Bretz et al. \cite{BretzPhilipE.1981Ioam}, who reported a stiffening effect and a reduced crack growth rate at water contents lower than $2.5$~wt.\% in nylon 66 samples. The so-called antiplasticization effect in glassy polymer systems was theorized some years later \cite{antiplas88}. \edits{In the recent literature however, the experimental evidence of the antiplasticization phenomenon in polyamides is rare \cite{Kehrer2024},} even if several works state that they provide data with low water contents \cite{fabre2018time}. The scarcity of experimental evidence could be attributed to the challenging control of low water contents in macroscopic PA66 samples and the current experimental scatter in the determination of $T_g$, which for some techniques can reach up to $10^o$C. 
It would be interesting to verify the MD predictions by experimentally performing high-resolution $T_g$ analyses and targeting the low water content regime.

$T_g$ marks the onset of a substantial motion of chain segments. This effect disrupts polymer packing and allows chain segments to reorient. One of the main relaxation modes of PA66 involves the localized motion of the hydrogen-bonded amide group ($\alpha$ and  $\beta$ relaxations) \cite{lukasheva2017influence}. Having access to the exact atomic positions of the distinct amide group, it would be insightful to observe the fluctuations of such group as $T$ and $C_w$ increase. The fluctuations of the O$_{amide}$ atoms were investigated using root mean square fluctuations RMSF ( $= \sqrt{< (\hat{r_i}-<\hat{r_i}>)^2>}$ ) around the atomic average position, where $r_i$ refers to the atomic position of atom $i$ (similar results are obtained with H$_{amide}$).

\begin{figure}
    \centering
    \includegraphics[height=0.55\textwidth]{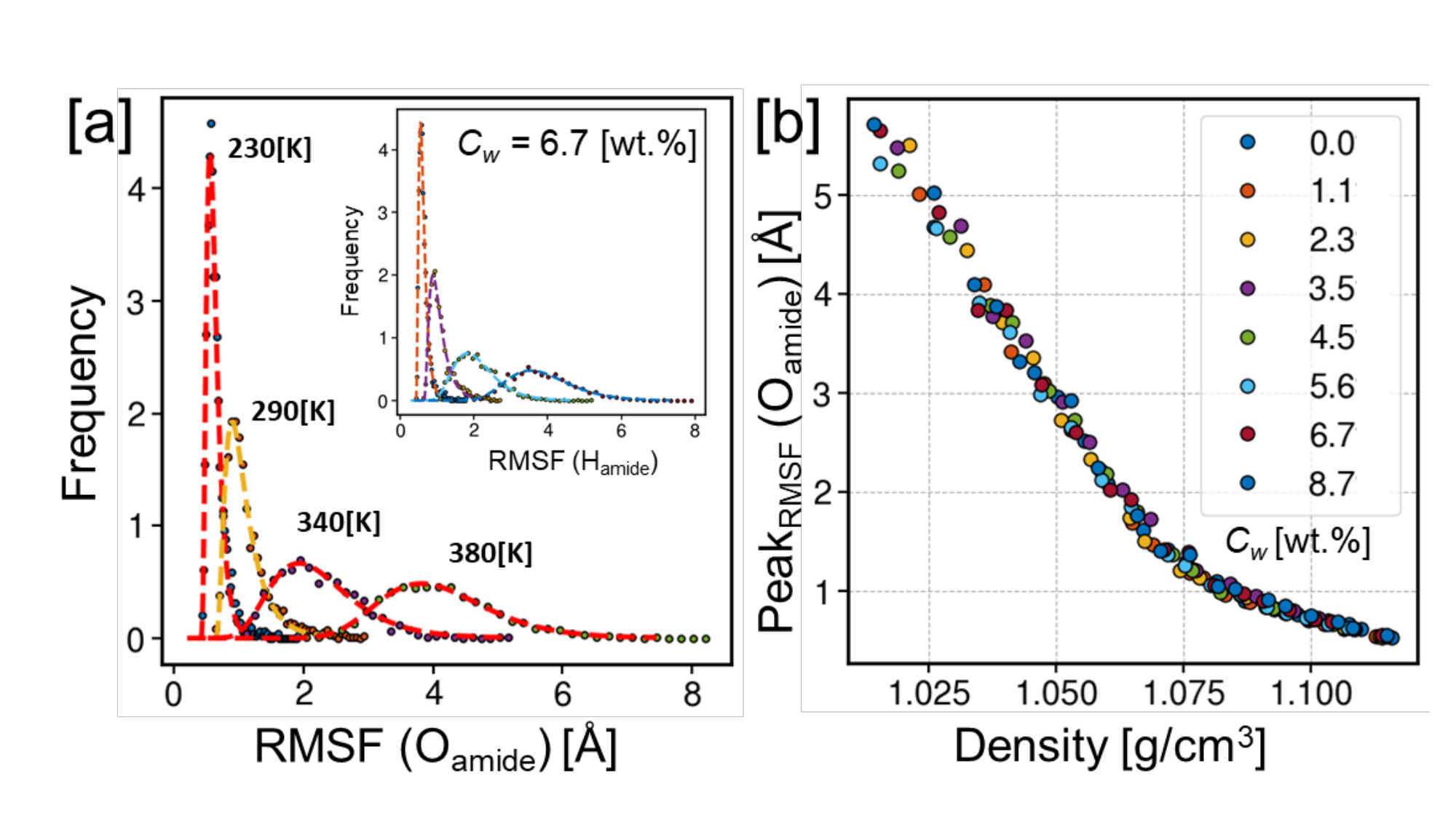}
    \caption{
Amide-group fluctuations in amorphous PA66 at different temperatures and water contents.
(a) Distribution of RMSF of amide oxygen atoms (O$_{amide}$) at $C_w= 6.7$~wt.\% obtained from MD simulations at various temperatures (Inset for H$_{amide}$). The MD data (symbols) are accurately described by log-normal fits (dashed lines) for all temperatures, reflecting the multiplicative nature of local molecular fluctuations. Increasing temperature shifts the RMSF distribution peak toward larger values and broadens its width, indicating enhanced segmental mobility and intensified uncorrelated motion of individual amide groups.
(b) Correlation between the peak of the RMSF (O$_{amide}$) distribution and $\rho(T)$ for all simulated temperatures and water contents. The data collapse onto a single master curve, revealing a universal coupling between local segmental dynamics and bulk packing density. A clear transition is observed at $\rho(T) \sim$  1.07 g/cm$^3$, separating the high-density, glassy regime from the low-density, rubbery regime associated with thermally or moisture-induced softening of the amorphous matrix. 
    }
    \label{fig:freq}

\end{figure}

Fig.~\ref{fig:freq}a shows the distribution of the RMSF of O$_{amide}$ for different temperatures (at $C_w = 6.7$~wt.\% as an example). At low temperatures, the O$_{amide}$ atoms exhibit a low magnitude of fluctuations, with a narrowly peaked distribution (only $\sim 0.5$~\AA~of RMSF at $230$~K), which indicates the rigid state of the overall amorphous PA66 and the confined vibrations of the functional groups. On the other hand, raising the temperature to $380$~K results in a $\sim$ 8X shift of the peak of the RMSF to $\sim$4~\AA. In addition, a significant broadening of the distribution is observed, which points to an uncorrelated movement of O$_{amide}$ atoms and a delocalized large-scale motion. The distribution of atomic fluctuations can be described by a log-normal distribution for the different simulation conditions (fit was conducted for all $T$ and $C_w$ – data not shown). Emergence of a log-normal distribution  
is a demonstration of the multiplicative {nature of the} random fluctuating events taking place in the polymer matrix \cite{chen2021local}.

The control over the system temperature is translated in MD simulations into a change in atomic velocity, which manifests itself in {the increase of fluctuations, the reduction in density, the observation of thermal expansion, etc. In that sense, we can}  
remove the explicit dependence on temperature and investigate the correlation between internal atomic fluctuations and system density. Fig. \ref{fig:freq}b plots the peak of the RMSF-O$_{amide}$ distribution for different {temperatures and water contents as a function of the system density.}  
Unsurprisingly, {as system density increases there is a reduction of atomic fluctuations.}  
In other words, the system undergoes expansion through the disruption of the polymer packing, {driven either by temperature or water absorption}. On the other hand, {temperature-water content superposition behavior in the amorphous PA66}  
emerges, where a master curve is {able to describe the packing state of the system for all the temperatures and water contents used in this work.}  
The master curve exhibits two linear regimes (low and high density) with a transition that takes place at $\rho$ $\sim 1.07$~g/cm$^3$. {This transition density can be achieved either by changing temperature or water content.}  
{The temperature-humidity equivalence in the mechanical response of PA systems has been reported elsewhere \cite{launay2013modelling, MIRI2009757} and has been attributed to the equivalent effects of temperature and water on the free volume of the PA66 amorphous phase. The same argument would be valid to explain the temperature-humidity equivalence on the dilatometric properties of amorphous PA66. This behavior also implies that thermal expansion and moisture expansion in unconfined amorphous PA66 are equivalent in nature.}   
It would be insightful to assess the prevalence of this duality in terms of thermal expansion and swelling for different polymer/solvent systems.

\subsection{Effect of water on the thermomechanical behavior of {amorphous} PA66}

{Water uptake has well-known consequences on the mechanical behavior of polyamides. The literature often refers to a plasticizing or softening effect, which leads to a decrease in stiffness or an increase in strain at break. Our goal in this work is to study this phenomenon via MD by focusing on the viscoelastic behavior of amorphous PA66 at different temperatures and water contents.} 

\begin{figure}
    \centering
    \includegraphics[height=0.55\textwidth]{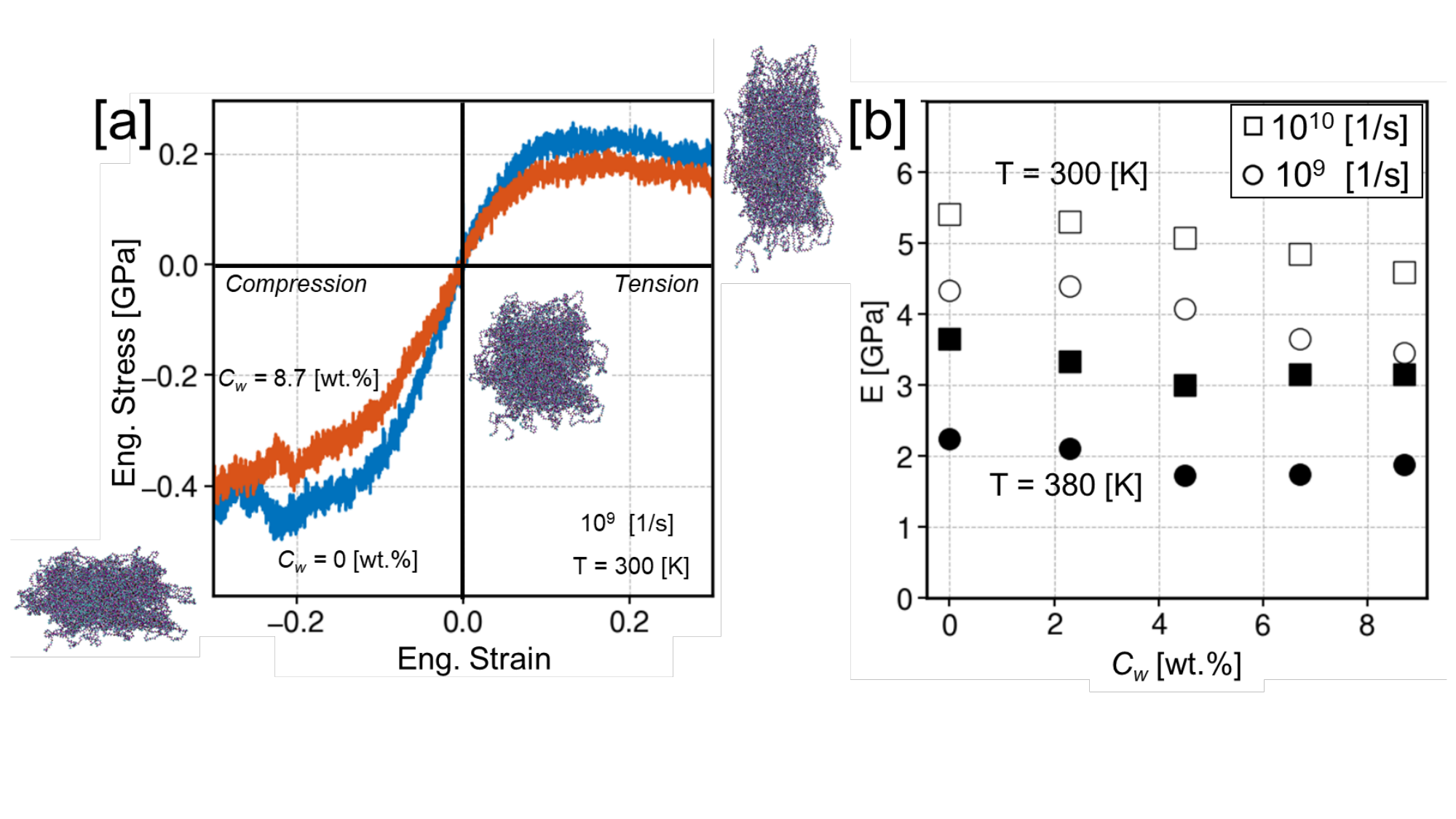}
    \caption{ 
Mechanical behavior of amorphous PA66 (dry and hydrated conditions) under uniaxial monotone loading.
(a) Engineering stress vs. strain curves from uniaxial deformation tests (300 K, $10^9$~1/s) in tension and compression mode. The introduction of water results in a reduction of both stiffness and yield strength, consistent with the plasticizing effect of absorbed moisture within the amorphous PA66 matrix.
(b) Young’s modulus (E) extracted from uniaxial tests at various temperatures, strain rates, and water contents. The modulus exhibits strong sensitivity to both temperature and deformation rate: an increase in temperature from 300 K to 380 K leads to a $\sim$35\% reduction in E for the dry polymer, while lowering the strain rate produces an additional decrease in stiffness, highlighting the viscoelastic character of amorphous PA66. Increasing water content induces an overall softening, particularly accentuated at low temperature (300 K).
    }
    \label{fig:ss}

\end{figure}

Fig.~\ref{fig:ss}a presents representative engineering stress vs. strain curves under both tension and compression at $300$~K and $10^9$~1/s, with concatenated data. {From a general perspective,} stress increases linearly with strain in the elastic region, followed by {a plastic deformation, which starts at a} yielding transition {and manifests a} chain sliding process. On the other hand, the elastic {regime appears to be larger } 
in compression than in tension. When comparing the dry system ($C_w = 0$~wt.\%) to the hydrated case ($C_w = 8.7$~wt.\%), we observe a reduction in both {Young's modulus} and yield strength (e.g., tensile yield strength drops from $230$~MPa to $180$ MPa), confirming the mechanical softening associated with water uptake.

Fig.~\ref{fig:ss}b summarizes the extracted Young’s modulus as a function of water content for two temperatures ($300$~K and $380$~K) and two strain rates ($10^9$ and $10^{10}$ 1/s). The temperature dependence is evident: Increasing the temperature from 300 K to 380 K reduces the Young's modulus by approximately 35\%. This result is consistent with the known thermal softening behavior of amorphous polymers {and reveals the activation of chain mobility in the rubbery state.} Furthermore, {we observe also a dependence on strain rate, revealing the viscoelastic nature of the system,} 
where higher strain rates result systematically in higher moduli; for instance, for dry amorphous PA66 at 300 K, the Young's modulus decreases from 5.4 GPa at 10$^{10}$ 1/s to 4.2 GPa at 10$^9$ 1/s.

We observe an overall softening effect in this range of water content (from dry to $8.7$~wt.\%), which is more prominent at low temperature ($300$~K). This larger drop in Young's modulus at $300$~K is probably the manifestation of a glass-to-rubbery transition induced by absorbed water. In addition, the drop in the modulus value is marginal in the range of water content between $0$ to $2.2$~wt.\%. This behavior could be attributed to the antiplasticizing effect discussed in the previous section and captured by the particular conditions of the MD uniaxial deformation test.

It is well established that the mechanical properties of amorphous polymers are dependent on both deformation rate and temperature. {This behavior is a clear manifestation of their viscoelastic nature. In practice, the same change of Young's modulus can be reached by either a certain increase in temperature or an equivalent reduction in strain rate.}
This {equivalence} relationship is the basis for the time-temperature superposition (TTS) principle, a widely employed method for constructing master curves {of a given mechanical quantity at different temperatures and frequencies in a reduced variable space.} In this work, {we build master curves of the Young's modulus as a function of strain rate (for a fixed water content) by sliding horizontally the data at different temperatures} using a temperature-dependent shift factor $a_T$. The shift factors are calculated relative to an arbitrary choice of a reference temperature $T_0$. 

\begin{figure}
    \centering
    \includegraphics[height=0.55\textwidth]{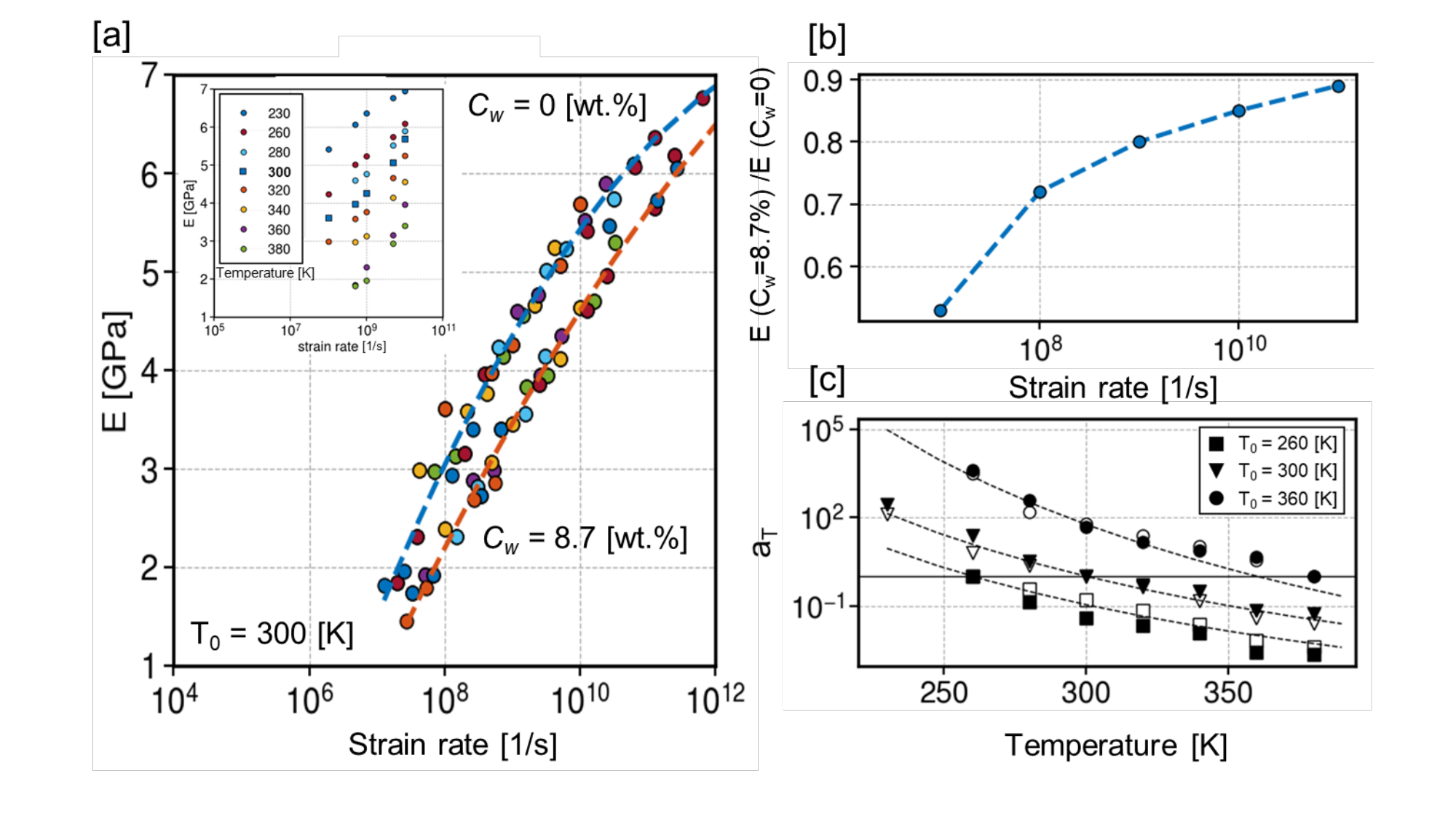}

    \caption{
Viscoelastic response of amorphous PA66 under dry and hydrated conditions analyzed using time-temperature superposition (TTS). (a) Master curves of Young’s modulus (E) as a function of strain rate for $C_w = 0$~wt.\% and $C_w = 8.7$~wt.\% using $300$ K as reference temperature. Data extracted at different absolute temperatures for a dry system are shown in the inset. The hydrated polymer curve is shifted to lower stiffness, indicating a reduced modulus at a given strain rate compared to the dry system (lines added for clarity).
(b) Strain-rate dependence of the hydration effect on E. At high strain rates, only a reduction in modulus of $\sim 10\%$ is observed upon hydration, whereas at low strain rates the same water uptake produces a reduction in modulus of almost 50\%. This behavior emphasizes the timescale-dependent nature of the plasticization process. 
(c) Temperature dependence of the TTS shift factor $a_T$ for different reference temperatures. The data  follow a classical Arrhenius-like trend and are well described by the Eyring equation (Eq.~\ref{eq:TTS}) in dashed lines (only shown for the dry polymer). Hydrated PA66 (filled symbols) shows little deviation from the dry system (open symbols), indicating comparable activation energies.
    }
    \label{fig:tts}

\end{figure}

Fig.~\ref{fig:tts}a shows {the obtained master curves of Young's modulus as function of strain rate (dry system and $C_w$ = 8.7 wt.\%)} using $T_0 = 300$ K.  
The narrow spread of the points confirms {indirectly the quality of the simulation results and} the time-temperature equivalency of the amorphous PA66 systems. {The relative position of the master curves manifests the softening effect driven by the water within the amorphous PA66. The similarity of both curves would allow an additional curve mastering based on the temperature-humidity equivalence. In fact, some literature points out that the mechanical behavior of PA systems under different hydro-thermal conditions follows a temperature-water content equivalence controlled by the temperature difference to the glass transition at a given water content $T - T_g(C_w)$ \cite{launay2013modelling,fabre2018time,polym15163387}. \edits{Taking into account this extended equivalence, some works have introduced horizontal and vertical shifts to construct master curves of measured viscoelastic properties accounting for time, temperature, and water content dependencies in polyamide materials \cite{Kehrer2023,Keursten2023}.}} While the gap between MD strain rates and typical experimental ones persists, these findings \edits{confirm in-silico the thermo-rheological behavior of water-polyamide systems and} provide insights on how to bring the MD results closer to what is observed experimentally. Fig.~\ref{fig:tts}b shows how the water-driven drop in stiffness depends on the strain rate. This result points to the underlying relaxation timescale needed for the internal structure of the hydrated PA66 matrix to respond to deformation.

Fig.~\ref{fig:tts}c {presents the variation of $a_T$ with temperature for three different reference temperatures $T_0$:} 260, 300 and 360~K. {$a_T$ is shown to be a decreasing function in the temperature range of this work: 230-380 K. Considering the $T_g$ estimations in §\ref{sec:results-Tg} (at cooling-rates of ~$10^{11}$~K/min), this mastering temperature range covers the glassy state, the glass transition and the rubbery flow regime. Given the extended range of mastering temperatures, we suppose that an Arrhenius-like equation is capable of describing the temperature dependence of $a_T$. In fact, Arrhenius-like behavior has been observed in low-temperature regions (from $T_g - 50$~K to $T_g$) and in the flow regime ($T >> T_g$) \cite{GSellChristian1994Iàlm}.}  
Concretely, to describe the temperature dependence of $a_t$ we propose using Eyring equation \cite{sahputra2013effects}: 

\begin{equation}\label{eq:TTS}
    \ln \left(a_{T}\right)=\frac{Q}{R}\left(\frac{1}{T}-\frac{1}{T_{0}}\right)
\end{equation}

where $R$ is the {universal} gas constant and $Q$ is the associated activation energy.  
{The fitting of each MD master curve (at a given $T_0$) by means of the Eq.~\ref{eq:TTS} is given in Fig.\ref{fig:tts}c.}

Only fits for dry polymer (unfilled markers) are shown for clarity. Filled markers correspond to amorphous PA66 with a water content $C_w = 8.7$ wt.\%. {The presence of water in the amorphous PA66 has little to no impact on the $a_T$ values compared to the dry system} and the observed differences appear to be within the error of the simulation results. {This result supports the temperature-water content equivalence and the possibility of getting a super master curve by collapsing the data at a reference water content.} {Using the MD data for the PA66 system, we identified an activation energy $Q = 50 \pm 12$ kJ/mol, independent of $C_w$}, which is slightly lower than the activation energy for viscous flow of nylon 66 ($70-76$ kJ/mol) \cite{shenoy1996thermoplastic,CecciaSimona2014Iopo}. The current findings \edits{offer interesting perspectives for further} research to accurately describe the {temperature dependence} of $a_T$ and to assess the applicability of different time-temperature superposition theories within different mastering temperature ranges \cite{sahputra2013effects,ljubic2014time}.   

\section{Conclusion}

This study used MD simulations to elucidate the mechanisms that govern the hygromechanical behavior of amorphous PA66 by varying the water content and  temperature. The simulations reproduced {well-known experimental behavior in polyamide systems,} including the depression of the glass transition temperature at large water contents and the reduction of elastic modulus with increasing moisture and temperature. 

{$T_g$ followed a non-monotonic dependence on water concentration, in which isolated water molecules seemed to stabilize the amide hydrogen-bond network, while water clustering beyond $\sim2.5$~wt.\% lead to a pronounced softening of the amorphous polymer matrix. This in silico result confirmed previous experimental evidence of the antiplasticization effect in polyamides with low water content.}

By quantifying amide-group fluctuations, we established a master relationship between local segmental mobility and overall amorphous polymer density, demonstrating a unified dependence on both temperature and humidity. This coupling provided a molecular foundation for the experimentally observed equivalence between thermal and moisture-induced {volumetric expansion.} 

{The MD calculated Young's modulus at different temperatures and strain rates exhibited time-temperature superposition for dry and hydrated systems. Considering the results in terms of volumetric expansion and stiffness, we revealed the molecular basis explaining the time-temperature-humidity equivalence in amorphous PA66. Given that this equivalence is also experimentally observed in semi-crystalline PA66 systems, we deduce that the amorphous phase plays a predominant role in this phenomenon.}

Future efforts should extend this MD framework to \edits{time-dependent protocols such as stress relaxation, creep, and recovery, performed systematically as a function of water content, and temperature. This would allow direct quantification of energy dissipation, recoverable versus permanent strain, and characteristic relaxation times, enabling a clearer distinction of the effect of water on the viscoelastic and viscoplastic behavior of amorphous polyamide. Furthermore, it is relevant to include the} semi-crystalline phase of PA66 systems to capture the coupling between crystalline and amorphous domains, where water preferentially interacts with the amorphous phase. {Accounting for the} crystalline phase {will constrain the system and MD simulations will require access to longer } 
relaxation phenomena to reproduce the full viscoelastic spectrum relevant to engineering applications.{The vision is to extract the relevant material parameters from atomistic simulations for }calibrating mesoscale or continuum constitutive models, which would 
provide a bridge between molecular dynamics and predictive finite element simulations for the virtual design of polyamide components.

\bibliographystyle{ieeetr}
\bibliography{fuelCell}

\end{document}